\documentclass[usenatbib]{mn2e}
\usepackage{epsfig}
\usepackage{psfig}
\usepackage{amsmath,amssymb}
\usepackage{txfonts}
\allowdisplaybreaks
\def\NMODY{{\sc n-mody\,}}
\def\FVFPS{{\sc fvfps\,}}

\newcommand\conc{c_{200}}
\renewcommand\d{{\rm d}}

\newcommand\fv{{f_{\rm v}}}
\newcommand\fHtwo{f_{\rm H_2}}

\newcommand\hz{h_z}
\newcommand\hzinst{h_{z,{\rm inst}}}
\newcommand\Hzero{H_0}

\newcommand\kms{{\rm \,km\,s^{-1}}}
\newcommand\kpc{{\rm \,kpc}}
\newcommand\K{{\rm \,K}}
\newcommand\kb{k_{\rm B}}

\newcommand\kinst{k_{\rm inst}}
\newcommand\kcrit{k_{\rm crit}}

\newcommand\kv{{\bf k}}

\renewcommand\mp{m_{\rm p}}
\newcommand\MDM{{M}_{\rm DM}}

\newcommand\Mgas{{M}_{\rm gas}}

\newcommand\Minst{M_{\rm inst}}
\newcommand\Mclump{M_{\rm clump}}
\newcommand\Mpc{{\rm \,Mpc}}

\newcommand\Msun{{M}_{\odot}}
\newcommand\Myr{{\rm \,Myr}}
\newcommand\Mzero{M_0}

\newcommand\nHtwo{n_{\rm H_2}}
\newcommand\ngas{n_{\rm gas}}
\newcommand\Nclump{{N}_{\rm clump}}
\newcommand\NDM{{N}_{\rm DM}}

\newcommand\pc{{\rm \,pc}}

\newcommand\rs{{r_{\rm s}}}
\newcommand\rvir{{r_{200}}}
\newcommand\rt{{r_{\rm t}}}
\newcommand\Rinst{{R_{\rm inst}}}

\newcommand\tcool{t_{\rm cool}}
\newcommand\tcross{t_{\rm cross}}
\newcommand\tdyn{t_{\rm dyn}}
\newcommand\tfric{t_{\rm fric}}
\newcommand\Tvir{T_{\rm vir}}

\newcommand\varepsilonclump{{\varepsilon_{\rm clump}}}

\newcommand\vR{{v_R}}

\newcommand\vphi{{v_{\phi}}}
\newcommand\vcirc{v_{\rm circ}}
\newcommand\vs{v_{\rm s}}

\newcommand\lambdainst{\lambda_{\rm inst}}
\newcommand\lambdacrit{\lambda_{\rm crit}}

\newcommand\rhocrit{\rho_{\rm crit}}
\newcommand\rhoDMzero{\rho_{\rm DM,0}}
\newcommand\rhoDM{\rho_{\rm DM}}
\newcommand\rhogas{\rho_{\rm gas}}
\newcommand\rhogaszero{\rho_{\rm gas,0}}

\newcommand\PhiDM{\Phi_{\rm DM}}
\newcommand\Phigas{\Phi_{\rm gas}}
\newcommand\Phieff{\Phi_{\rm eff}}
\newcommand\Phieffzero{\Phi_{\rm eff,0}}

\newcommand\Sigmagas{\Sigma_{\rm gas}}
 \let\K=\kelvin
\renewcommand\L{\mathcal{L}}

\newcommand\yr{{\rm \,yr}}
\begin{document}
%\begin{landscape}

\date{Accepted 2014 October 22.  Received 2014 October 3; in original form 2014 June 17}

\title[Dark matter in dwarf spheroidal galaxies]{Early flattening of dark matter cusps in dwarf spheroidal galaxies}
\author[C. Nipoti and J. Binney]{Carlo Nipoti$^{1}$\thanks{E-mail: carlo.nipoti@unibo.it} and James Binney$^{2}$ 
\\ $^{1}$ Department of Physics and Astronomy, Bologna University,
viale Berti-Pichat 6/2, I-40127 Bologna, Italy
\\ $^{2}$ Rudolf Peierls Centre for Theoretical Physics, Keble Road,
Oxford OX1 3NP, UK }

\maketitle
\begin{abstract}
Simulations of the clustering of cold dark matter yield dark-matter
halos that have central density cusps, but observations of totally
dark-matter dominated dwarf spheroidal galaxies imply that they do not
have cuspy central density profiles. We use analytic calculations and
numerical modelling to argue that whenever stars form, central density
cusps are likely to be erased.  Gas that accumulates in the potential
well of an initially cuspy dark-matter halo settles into a
disc. Eventually the surface density of the gas exceeds the threshold
for fragmentation into self-gravitating clouds.  The clouds are
massive enough to transfer energy to the dark-matter particles via
dynamical friction on a short time-scale.  The halo's central cusp is
heated to form a core with central logarithmic density slope
$\gamma\approx 0$ before stellar feedback makes its impact. Since star
formation is an inefficient process, the clouds are disrupted by
feedback when only a small fraction of their mass has been converted
to stars, and the dark matter dominates the final mass distribution.
\end{abstract}

\section{Introduction}

The observed stellar velocity dispersions of dwarf spheroidal galaxies
(dSphs) cannot be explained by assuming that the stars are in
equilibrium in a gravitational potential dominated by the baryons: the
standard interpretation of the observed stellar kinematics is that the
mass distribution of dSphs is everywhere dominated by dark matter
(DM), with mass to light-ratios $M/L\approx 10-100 (M/L)_{\odot}$
\citep{Aar83,Fab83,Mat98}.  This property makes dSphs natural
laboratories to study the density distribution of DM halos.  However,
although most observational studies agree that DM dominates over
baryons even within the half-light radius
\citep[e.g.][]{Str08,Wal09,Bat13,Col14}, it is still debated whether
the inner distribution of DM is characterized by a cusp or a core
\citep[e.g.][]{Bre13,Jar13}. {A consensus is emerging that the best
observations disfavour the presence of steep central DM cusps such as
those predicted by cosmological $N$-body simulations in a concordance
$\Lambda$ cold dark matter universe \citep{Bat08,Wal11,Agn12,Amo13}.}

Our understanding of the process of formation of dSphs is still far
from complete. In the most popular scenario dSphs formed originally as
gas rich dwarf irregular or disc galaxies and then lost their gas as a
consequence of internal or environmental effects \citep[e.g.][and
references therein]{Lin83,May01,May06,Mil14}. A successful formation
model should be able to account for the all the properties of dSphs
that can be inferred from observations, such as the structure and
kinematics of the stellar distribution, metallicity of stars, and star
formation history. Several studies have been carried out trying to
reproduce at least some of these properties with either cosmological
hydrodynamical simulations \citep{Saw10,Saw12,Sim13}, simulations of
isolated halos \citep{Rev09,Ass13} or semi-analytic models
\citep{Li10,Fon11,Rom13,Sta13}. Here we focus on the two specific
properties of dSphs: the facts that DM dominates the mass budget
throughout these galaxies and that the present-day DM profile is not
divergent in the central regions.  

Several mechanisms are likely to jointly determine the DM
distribution of dSphs: baryon removal by supernova feedback
or ram pressure \citep[]{Nav96,Rea05,Zol12,Arr14}, clumpy baryonic
infall \citep*{Elz01,Goe10,Col11} and stellar-feedback driven fluctuations
of the gravitational potential \citep*{Mas06,Mas08}.  This last
mechanism has recently drawn the attention of several authors who
explored its feasibility within a cosmological framework
\citep{Gov12,Gar13,Amo14,Mad14}. However, given the uncertainties on
the physics of stellar feedback, it is unclear whether the properties
of present-day dSphs can be explained by the associated gravitational
potential fluctuations \citep{Pen12,Pon14}.  Here we argue that in the
progenitors of dSphs the distribution of DM was substantially affected
by baryons {\it before} star formation got underway, because in these
systems star formation is inevitably preceded by the formation of
massive baryonic clumps.  The orbital decay of these clumps erased the
central cusps of their host halos.

We present a coherent picture that leads from the cosmological DM halo
to the matter distribution of a present-day dSph, taking as reference
objects the classical dSphs of the Milky-Way. We do not attempt to
simulate self-consistently the entire formation and evolution process,
but we focus on the most important steps, demonstrating quantitatively
that the proposed picture is feasible.

We envisage the following scenario for the formation
and early evolution of the baryonic and dark components of dSphs.
\begin{enumerate}
\item The host DM halo is initially cuspy as predicted by cosmological
  $N$-body simulations.
\item Gas accumulates in the halo potential well and settles into a disc.
\item The infall of gas initially tends to make the halo cuspier,
  but the effect is small as long as the gas density is lower
  than the DM density.
\item Once the gas density becomes higher than the local DM
  density, the gas disc becomes unstable and starts to
  fragment.
\item Since the gas temperature is not negligible with respect to the
  system virial temperature, the disc fragments into large clumps of
  gas.
\item These big clumps transfer energy to the DM particles via
  dynamical friction: the halo is quickly heated and the central cusp
  is flattened into a core.
\item {Subsequently, stars start to form in the clumps, and  efficient
  stellar feedback then expels most of the gas, so the final stellar
  density is much lower than both the initial gas density and the final
  DM density.} 
\end{enumerate}
The aim of this work is to give a quantitative description of the main
physical processes involved in the above scenario, focusing in
particular on demonstrating that fragmentation occurs where baryons
are dominant (iv), that the resulting gas clumps are very massive (v),
and that these clumps quickly flatten the original DM cusp (vi).  As
mentioned above, the idea that the DM profiles of dSphs are flattened
by baryonic clumps has been considered in other works
\citep[][]{Elz01,Goe10,Col11}. Here we put this proposal into a
broader context by modelling the coevolution of DM and baryons in
dSphs starting from cosmic gaseous accretion and estimating
self-consistently the properties of the gaseous clumps responsible for
the flattening of the DM density profile.

The paper is organized as follows. In Section~\ref{sec:mod} we define
the properties of the protogalaxy model. Section~\ref{sec:evol}
discusses the coevolution of baryons and DM.  In Section~\ref{sec:rel}
we relate our work to previous work on the subject.
Section~\ref{sec:con} concludes.

\section{Protogalaxy model}
\label{sec:mod}

\subsection{Dark matter halo}
\label{sec:dm}

We assume that the DM halo is represented initially by a spherical
\citet[][NFW]{Nav95} distribution with density
\begin{equation}
 \rhoDM(r) =  \frac{\rhoDMzero}{(r/\rs)(1+r/\rs)^2}
\end{equation}
and gravitational potential
\begin{equation}
\label{eq:phidm}
 \PhiDM(r) =  -4\pi G\rhoDMzero \rs^2\frac{\ln(1+r/\rs)}{r/\rs},
\end{equation}
where
\begin{equation}
 \rhoDMzero\equiv\dfrac{200}{3}\dfrac{\rhocrit\conc^3}{\ln(1+\conc)-\conc/(1+\conc)}.
\end{equation}
Here $\rs$ is the scale radius, $\conc\equiv \rs/\rvir$ is the
concentration and $\rvir$ is the virial radius. The critical density
of the Universe $\rhocrit(z)={3H^2(z)}/{8\pi G}$ depends on redshift
$z$ through the Hubble parameter
\begin{equation}
H(z)=\Hzero\left[\Omega_{m,0}(1+z)^3+\Omega_{\Lambda,0}\right]^{1/2},
\end{equation}
where $\Hzero$ is the Hubble constant, and $\Omega_{\rm m,0}$ and
$\Omega_{\Lambda,0}$ are the present-day normalized densities of matter and
dark energy. We assume $\Hzero=70\kms \Mpc^{-1}$, $\Omega_{\Lambda,0}=0.73$
and $\Omega_{\rm m,0}=0.27$.

In this work we focus on classical dSphs, such as the Milky-Way
satellites Fornax and Scupltor, having DM halos with masses of the
order of $10^9\Msun$.  These objects likely formed in lower density
environments at $z\sim 2$, because dwarf galaxies that formed at much
higher redshift will have done so in dense environments, and merged
into more massive systems by $z\sim 0$.  Given a halo mass and
redshift, the concentration $\conc$ can be estimated from cosmological
$N$-body simulations (e.g. \citealt{Mun11,Pra12,Ogi14,Ish13}). For our
reference halo model we fix $z=2$, $\MDM=10^9\Msun$, $\conc=5.69$, so
the virial radius is $\rvir\simeq10.3\kpc$ and the scale radius is
$\rs\simeq1.81\kpc$.

\subsection{Gas temperature}
\label{sec:temp}

In the first stages of galaxy formation gas will fall into
the halo potential well. In the considered case the halo is located in
a low-density environment, so the infalling gas should be either
primordial or very metal poor. It is often assumed that such gas is
shock heated to the halo virial temperature
\begin{equation}
\Tvir\equiv\frac{\mu\mp}{2\kb}\frac{G\MDM}{\rvir},
\end{equation}
where $\mu$ is the mean mass per particle in units of the proton mass
$\mp$. {For our reference halo ($\MDM=10^9\Msun$, $z=2$) we get
$\Tvir\simeq1.49\times 10^4\K$ with $\mu=0.6$,  but in practice most of the
gas is gas is less strongly heated than this \citep{Bin77,Bir03}, so
$10^4\K$ is a conservative upper limit on the temperature of the gas.} The ability of gas to cool below $10^4\K$
crucially depends on the gas metallicity, on the fraction of molecular
hydrogen and on the presence of a background radiation field (see,
e.g., \citealt{Sti09}, \citealt{Loe13} and reference therein). Rather
than attempting to model in detail the cooling process below
$T=10^4\K$, we consider two representative cases: one in which cooling
is more effective, and the gas temperature is $T=3\times 10^2\K$
(model T2) and the other in which cooling is less effective and the
gas temperature is $T=3\times 10^3\K$ (model T3).

%%%%%%%%%%%%%%FIG 1
\begin{figure*}
\centerline{\psfig{file=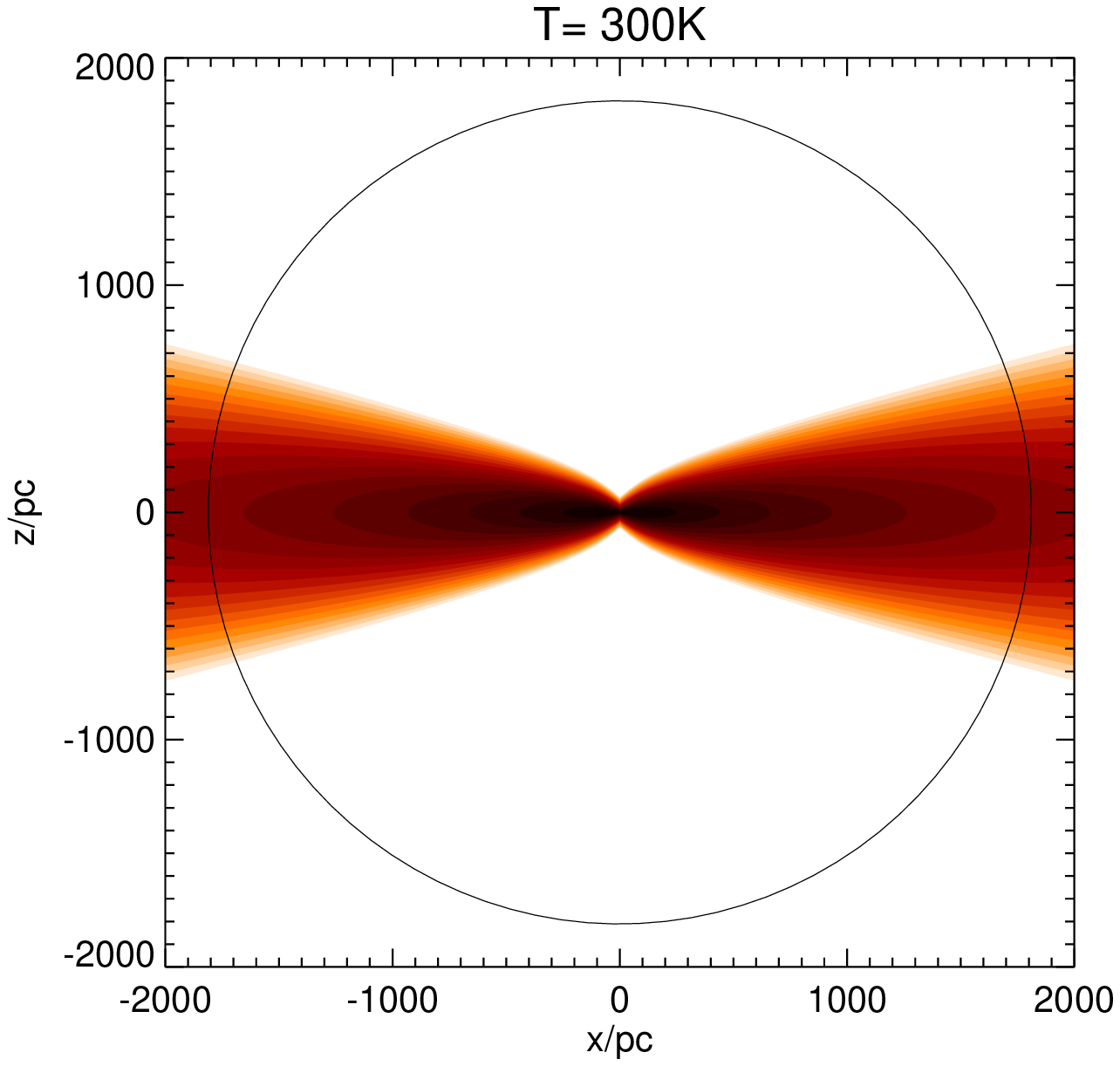,width=0.5\hsize,bbllx=0bp,bblly=0bp,bburx=400bp,bbury=361bp,clip=}\psfig{file=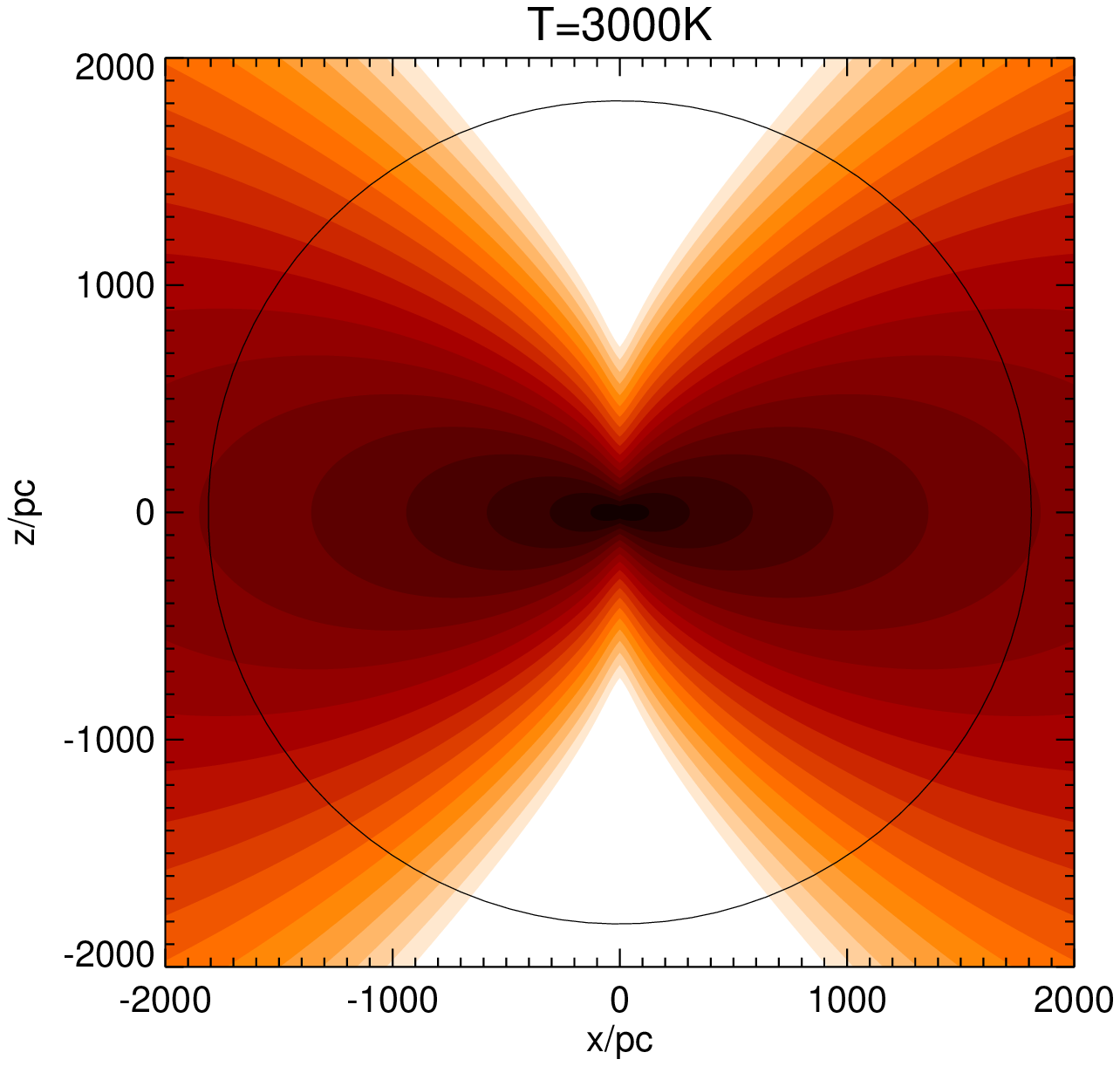,width=0.5\hsize,bbllx=0bp,bblly=0bp,bburx=400bp,bbury=361bp,clip=}}
\caption{Gas iso-density contours (logarithmically spaced by 0.5 dex)
  in the meridional plane of models T2 (left-hand panel) and T3
  (right-hand panel). Contours corresponding to densities lower than
  the maximum gas density by more than 10 orders of magnitude are not
  shown.  In each panel the circle indicates the DM halo scale radius
  $\rs$.}
\label{fig:map}
\end{figure*}
%%%%%%%%%%%%%%%%%%%%%%%

\begin{table*}
%\begin{center}
\caption{Parameters of the models T2 and T3.  $T$: gas temperature.
  $\Mgas$: total gas mass. $\fv\equiv\Omega R/\vcirc$: normalization
  of the rotation law. $\lambdainst$: wavelength of the
  fastest-growing mode (equation~\ref{eq:lambdainst}). $\Rinst$:
  radius at which the instability occurs.  $\hzinst$: vertical
  scale-height at $\Rinst$. $\Minst$: mass corresponding to the
  fastest-growing unstable mode (equation~\ref{eq:Minst}). The DM halo
  is the same in both models (see Section~\ref{sec:dm}).
\label{tab:par}}
\begin{tabular}{lllllllll}
\hline Model & $T/K$ & $\Mgas/\Msun$ & $\fv$ & $\lambdainst/\pc$ &
$\Rinst/\pc$ & $\lambdainst/2\pi\Rinst$ & $\lambdainst/\hzinst$ &
$\Minst/\Msun$ \\
%Name  & $T/K$  &  $\Mgas/\Msun$ & $\fv=\Omega R/\vcirc$ & $\lambdainst/\pc$ 
% &  $\Rinst/\pc$ &   $\lambdainst/2\pi\Rinst$  &  $\lambdainst/\hzinst$ & $\Sigma(0)/\Msun\pc^{-2}$ 
\hline
T2 & $3\times10^2$ & $1.168\times10^7$ & $0.9899$ & $124.9$ & $183.6$ & $0.1083$ & $7.275$ & $1.541\times10^5$ \\
T3 & $3\times10^3$ & $7.613\times10^7$ & $0.8938$ & $484.0$ & $240.5$ & $0.3202$ & $8.027$ & $6.048\times10^6$ \\
%T2 &  0.3000E+03 &   0.1168E+08 &   0.9899E+00 &   0.1249E+03 &   0.1836E+03 &   0.1083E+00 &   0.7275E+01 &     0.1541E+06\\
%T3 &  0.3000E+04 &   0.7613E+08 &   0.8938E+00 &   0.4840E+03 &   0.2405E+03 &   0.3202E+00 &   0.8027E+01 &     0.6048E+07\\
% T2 & $2\times10^2$ & $2.47\times10^6$ &  0.97 & 98  & 169 &  0.09 &  7.3 & 3.0 \\
% T3 & $2\times10^3$ &  $2.57\times10^7$ &  0.90 & 264 & 137 &  0.31 &  7.7 & 48.2 \\
\hline
\end{tabular}
%\end{center}
\end{table*}

%  200       1e+09   2.465e+06         999       98.26         169     0.09254       7.377       2.973
% 2000       1e+09   2.233e+07         999       263.6       136.7      0.3069       7.719        48.2

\subsection{Disc instability}
\label{sec:inst}

In the presence of angular momentum the infalling and cooling gas is
expected to settle into a disc. At low gas density the disc is stable,
but above a critical density the disc becomes gravitationally unstable
and fragments. We are therefore interested in determining the physical
conditions for the onset of such instability. For infinitesimally thin
gaseous disks such conditions are determined by Toomre's stability
criterion \citep{Saf60,Too64}.  A priori it is not obvious that the
thin-disc approximation is appropriate in the considered case. If the
gas temperature $T$ is much lower than $\Tvir$ the disc is thin, but
the thickness of the disc is not negligible if $T$ is a significant
fraction of $\Tvir$.  In any case, we proceed under the assumption
that the thin disc approximation is justified and we verify and
discuss this hypothesis a posteriori.

Here we briefly recall the relevant equations for a local stability
analysis of a self-gravitating gaseous disc \citep[e.g.][]{BT08}. The
dispersion relation for a perturbation with wave-vector $\kv$ is
\begin{equation}
\omega^2=\kappa^2-2\pi G \Sigmagas k +\vs^2 k^2,
\end{equation} 
where $\omega$ is the perturbation frequency, $k=|\kv|$ is the
perturbation wave-number, $\Sigmagas$ is the local gas surface
density, $\kappa$ is the epicycle frequency and $\vs$ is the adiabatic
sound speed.  The condition for disc stability ($\omega^2>0$ for all
$\kv$) can be written as $Q>1$, where
\begin{equation}
\label{eq:q}
Q\equiv\frac {\vs \kappa}{\pi G \Sigmagas}
\end{equation} 
is Toomre's parameter.  It can be shown that the same stability
criterion holds also when the gas coexists with other matter, such as
the DM halo: in this case the epicycle frequency $\kappa$ is larger
than that that generated by the gas on its own, and $Q$ is higher at
given gas surface density $\Sigmagas$.  When $Q<1$ the fastest growing
mode is the one with $k=\kinst$, where $\kinst$ is such that
$\d\omega^2/\d k=0$, so $\kinst={\pi
  G\Sigmagas}/{\vs^2}={2}\kcrit/{Q^2}$, where
$\kcrit\equiv{\kappa^2}/{2\pi G\Sigmagas}$ is the minimum unstable
wave-number. The wavelength of the fastest-growing unstable mode is
\begin{equation}
\label{eq:lambdainst} 
\lambdainst=\frac{2\pi}{\kinst}=\frac{2\vs^2}{G\Sigmagas}=\frac{2\pi^2 Q^2G\Sigmagas}{\kappa^2}=\frac{Q^2}{2}\lambdacrit,
\end{equation}
where $\lambdacrit\equiv{4\pi^2 G\Sigmagas}/{\kappa^2}$ is the longest
unstable wavelength. The mass corresponding to the fastest-growing
unstable mode is
\begin{equation}
\label{eq:Minst} 
\Minst= \pi\left(\frac{\lambdainst}{2}\right)^2\Sigmagas(\Rinst),
\end{equation}
where $\Sigmagas(\Rinst)$ is the gas density evaluated at the radius
at which the instability occurs.

%%%%%%%%%%%%%%FIG 2
\begin{figure*}
\centerline{\psfig{file=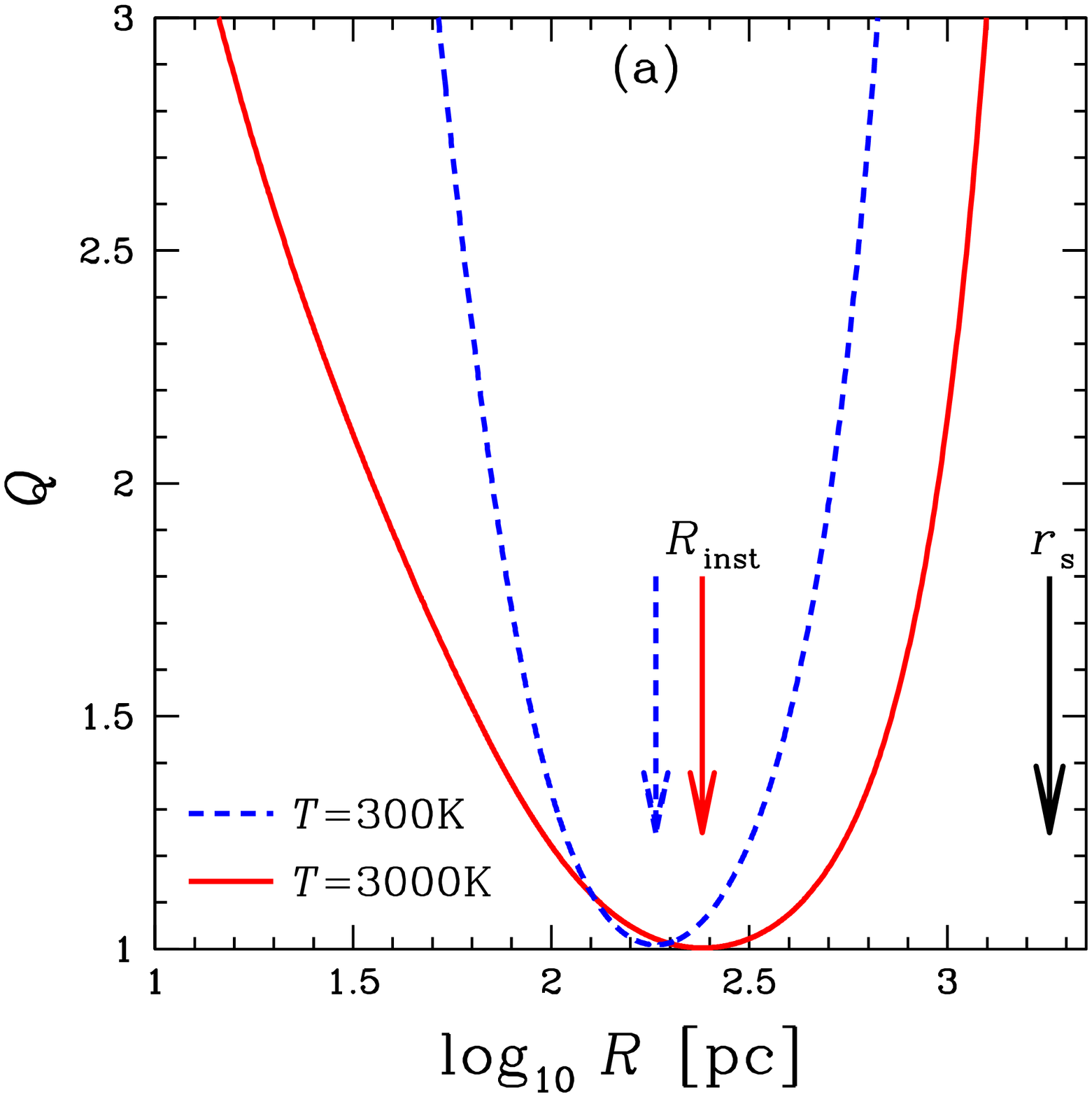,width=0.333\hsize}\psfig{file=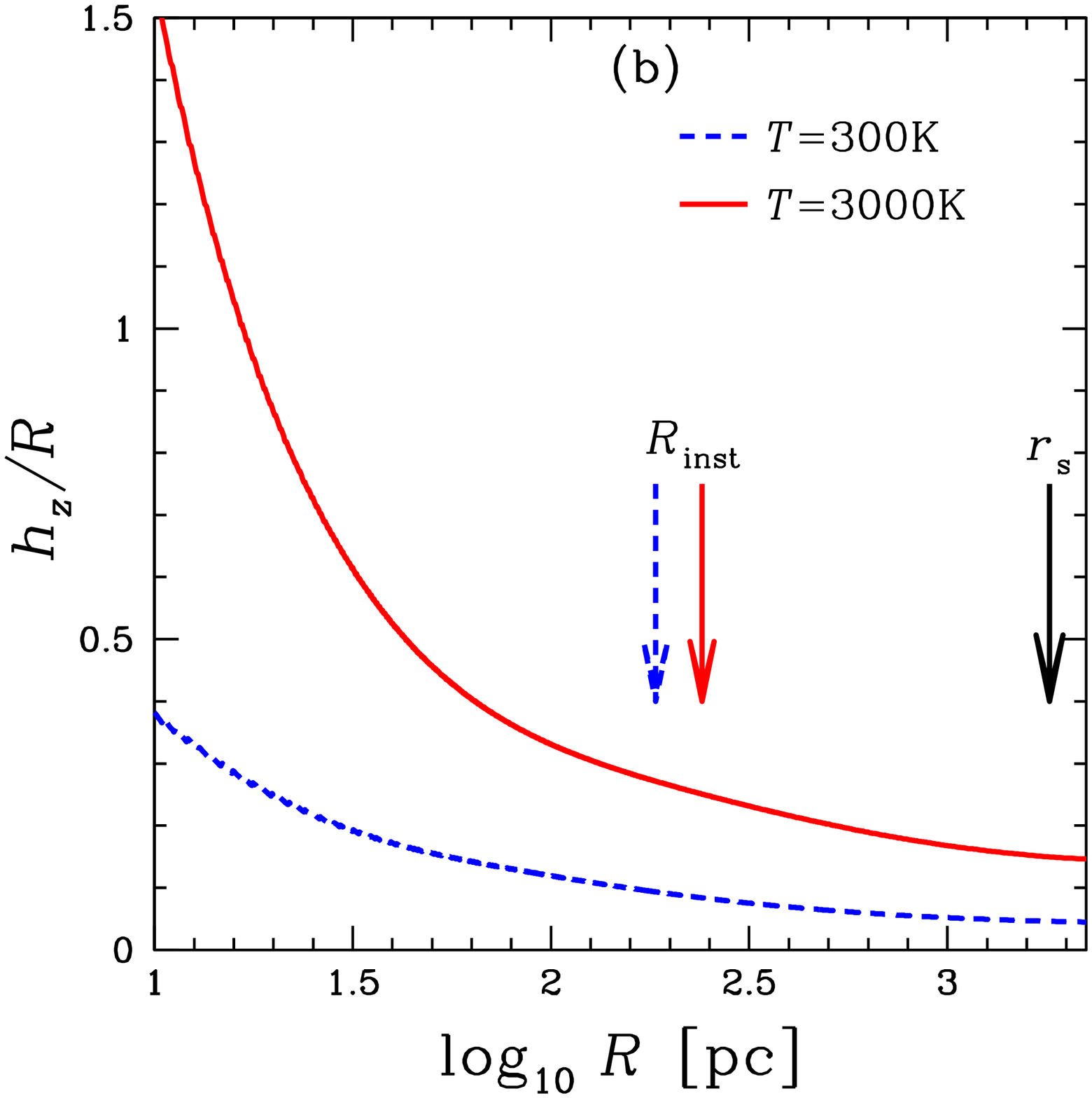,width=0.333\hsize}\psfig{file=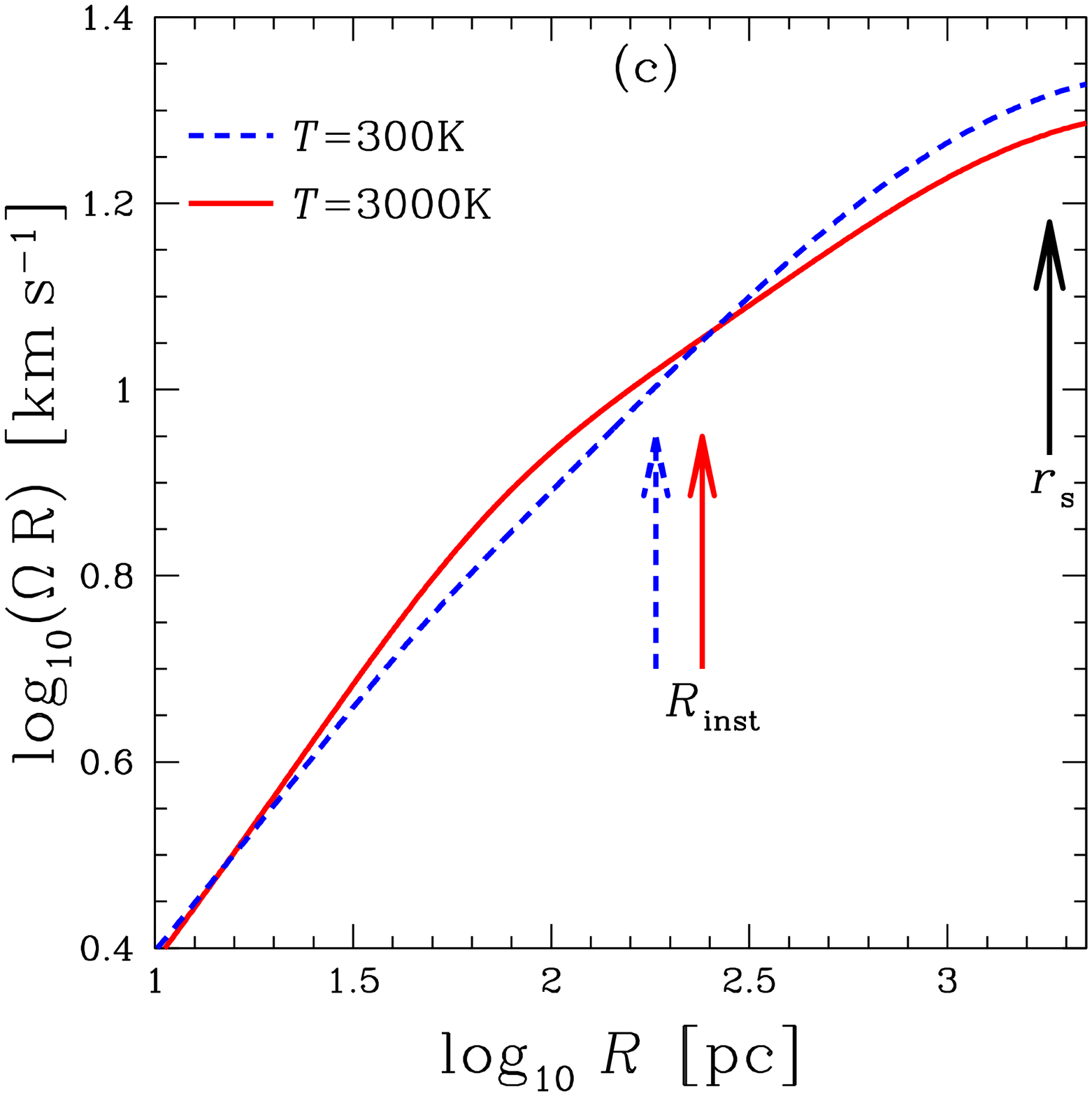,width=0.333\hsize}}
\centerline{\psfig{file=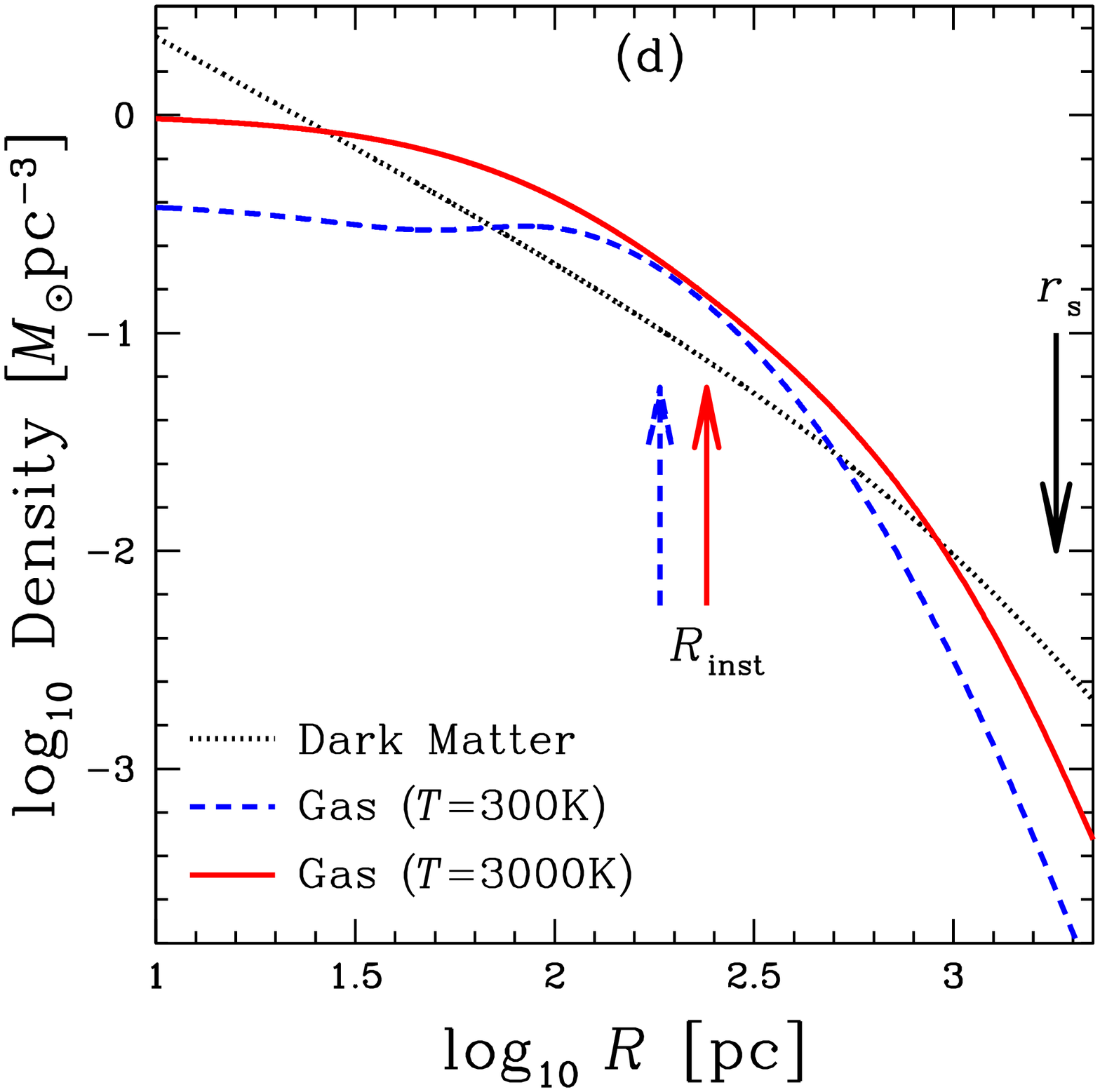,width=0.333\hsize}\psfig{file=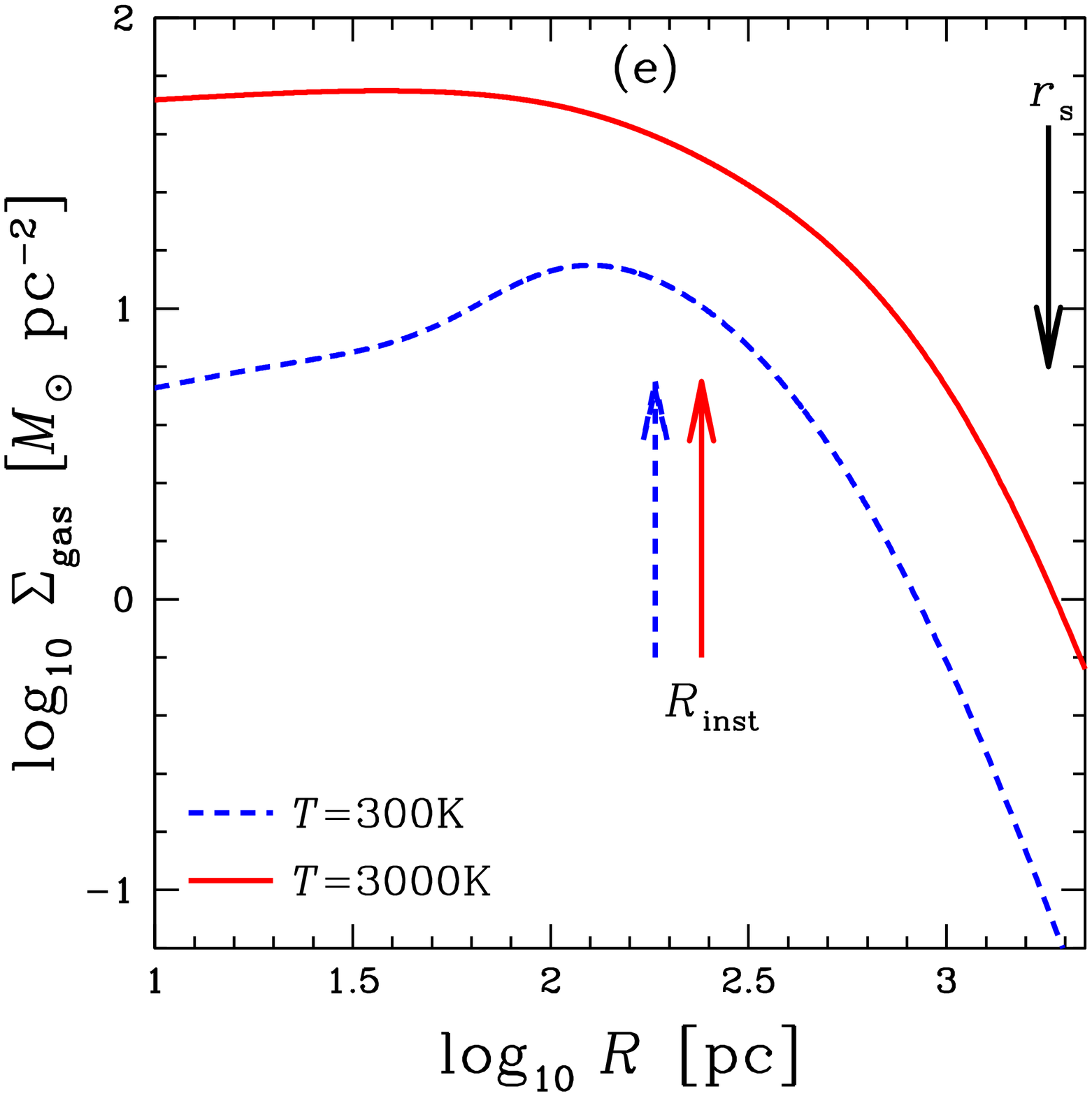,width=0.333\hsize}\psfig{file=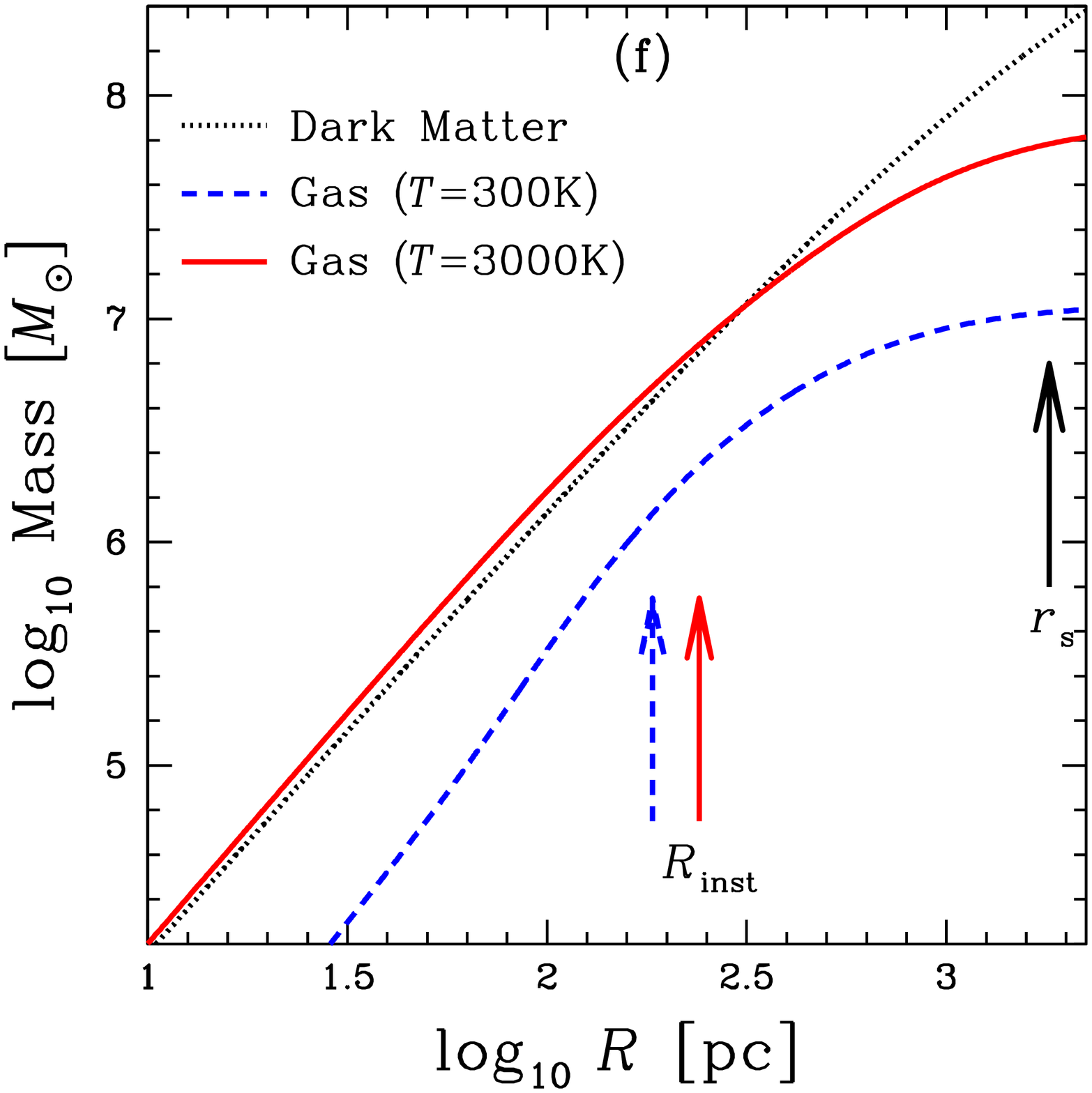,width=0.333\hsize}}
\caption{Radial profiles of $Q$ parameter (panel a), vertical
  scale-height $\hz$ (panel b), gas rotation speed (panel c), gas
  density at $z=0$ (panel c), gas surface density (panel e) and gas
  mass (panel f) for models T2 ($T=300\K$; dashed curves) and T3
  ($T=3000\K$; solid curves). In panels (d) and (f) the density and
  mass profiles of the DM halo are shown for comparison (dotted
  curves). The arrows indicate the gas instability radius $\Rinst$ and
  the halo scale radius $\rs$.}
\label{fig:all}
\end{figure*}
%%%%%%%%%%%%%%%%%%%%%%%

\subsection{Gas distribution}

Here we build three-dimensional axisymmetric disc models by computing
an equilibrium solution for a body of isothermal gas that rotates
differentially with angular velocity $\Omega$ in the total
gravitational potential $\Phi=\PhiDM+\Phigas$, where $\PhiDM$
(equation~\ref{eq:phidm}) and $\Phigas$ are the gravitational
potentials of DM and gas, respectively.  We work in cylindrical
coordinates $(R,z)$.  The momentum equation for isothermal gas at
temperature $T$ can be written
\begin{equation}
\frac{\kb T}{\mu \mp}\frac{\nabla\rhogas}{\rhogas}=-\nabla\Phieff,
\end{equation}
where $\rhogas$ is the gas density and 
\begin{equation}
 \Phieff(R,z)= \Phi(R,z) - \int^R\d R'\,R' \Omega^2(R')
\end{equation}
is the effective potential.   It follows that the
density distribution is given by
\begin{equation}
\displaystyle
\rhogas(R,z)\equiv \rhogaszero 
\exp\Bigg\{{-{\mu \mp\over\kb T}\left[\Phieff(R,z)-\Phieffzero\right]}\Bigg\},
\label{eq:rhogas}
\end{equation}
where $\rhogaszero$ and $\Phieffzero$ are the density and effective
potential at a reference point.  We assume rotation law
\begin{equation}
\Omega(R)=\fv{\vcirc(R)\over R},
\end{equation}
where $\fv<1$ is a dimensionless factor and $\vcirc$ is the circular
speed of the total (DM plus gas) distribution measured in the
equatorial plane, which is defined by
\begin{equation}
\frac{\vcirc^2}{R}=\frac{\d\Phi(R,0)}{\d R},
\label{eq:vcirc}
\end{equation}
where $\Phi(R,z)$ is the total gravitational potential in cylindrical
coordinates.  The spherical halo potential is given analytically by
equation~(\ref{eq:phidm}).  The gravitational potential of the gaseous
disc $\Phigas$ is computed numerically using the spherical
harmonics-based solver of the Poisson equation contained in the
numerical code \NMODY\ \citep*[][]{Cio06,Lon09}.  As the gas
distribution and velocity field depend on the gas potential and
vice-versa, we find a solution iteratively computing $\Phigas$,
$\rhogas$ and $\vcirc$ up to convergence.

It is useful to define the face-on surface density of the disc
\begin{equation}
\label{eq:sigma}
\Sigmagas(R)=2\int_0^{\infty}\d z\,\rhogas(R,z)
\end{equation}
and the vertical scale height of the distribution $\hz(R)$ such that
\begin{equation}
\int_R^{R+\Delta R} \d R' \int_0^{\hz} \d z'\, \rhogas(R',z')
=\frac{1}{2}\int_R^{R+\Delta R} \d R'  \int_0^\infty\d z'\, \rhogas(R',z'),
\end{equation}
i.e., $\hz(R)$ is a measure of the thickness at radius $R$ containing
half of the mass in a cylindrical shell of radius $R$
and width $\Delta R \ll R$.

For our three-dimensional disc models, the parameter $Q$
(equation~\ref{eq:q}) can be computed at any $R$ by combining $\Sigmagas(R)$
as given by equation~(\ref{eq:sigma}), the local epicyclic frequency
$\kappa(R)$, defined by
\begin{equation}
\kappa^2(R)=4\Omega^2+2\Omega R \frac{\d \Omega}{\d R},
\end{equation}
and the adiabatic sound speed $\vs=5\kb T/3\mu\mp$, which is
independent of radius for the considered isothermal models.

\subsection{Marginally stable disc models}
\label{sec:marg}

We consider a scenario in which the gas accumulates from low densities
(high $Q$) up to values of $\Sigmagas$ such that the $Q$-parameter
reaches the value $Q=1$ at some radius $\Rinst$, which is the radius
at which the the fastest instability occurs.  This fixes the gas
surface density normalization and thus $\rhogaszero$ appearing in
equation~(\ref{eq:rhogas}).  These ``marginally stable'' models are
representative of the physical conditions of gaseous discs just before
they fragment.

Here we present two representative marginally stable isothermal disc
models in equilibrium in the common potential of the disc itself and
of the NFW halo described in Section~\ref{sec:dm}: model T2
($T=3\times10^2\K$) and model T3 ($T=3\times10^3\K$). Here we adopt
$\mu=1.2$, as appropriate for primordial or very low-metallicity gas,
which is mostly neutral at the considered temperatures and densities
\citep[e.g.][]{Loe13}. For given gas temperature, we assume
$\fv=\sqrt{1-T/\Tvir}$, so $\fv\approx 1$ (maximum rotation) for
$T\ll\Tvir$ and $\fv\approx 0$ (no rotation) for $T\approx\Tvir$.  The
values of the parameters for models T2 and T3 are reported in
Table~\ref{tab:par}.

Figure~\ref{fig:map} shows the maps of the density distributions in
the meridional plane for models T2 and T3. In both cases the density
distribution is toroidal: the density contours appear like a flaring
disc when the temperature is lower (model T2) and tend to be
peanut-shaped at higher temperatures (model T3).  In
Fig.~\ref{fig:all} we show, for the same two models, the radial
profile of the $Q$ parameter, the ratio $\hz/R$ that measures the
thickness of the disc, the rotation speed $\Omega R$, the densities of
gas and DM, the surface density of gas $\Sigmagas$, and the masses of
gas and DM $M(R)$.  By construction $Q\geq1$ and $Q\simeq 1$ only at
$\Rinst$ (Fig. \ref{fig:all}a).  The values of $\hz/R$ suggest that
the disc can be considered relatively thin over most of the gas mass
distribution, and in particular $\hz/R\lesssim0.3$ at $\Rinst$
(Fig. \ref{fig:all}b). The shape of the speed profiles
(Fig. \ref{fig:all}c) are similar to the circular speed profile of the
underlying NFW DM halo, with only slight differences due to the
gravitational potential of the gas: the speed varies approximately as
$R^{0.5}$ at small radii but flattens at larger radii.  The profiles
of $\rhogas$ and $\rhoDM$ support the idea that the condition
$\rhogas>\rhoDM$ is required for fragmentation
(Fig. \ref{fig:all}d). The corresponding critical surface density for
instability is $10-40\Msun\pc^{-2}$ (Fig. \ref{fig:all}e).  The total
gas mass contained within $R=\rvir$ is $\Mgas\simeq 1.2\times
10^7\Msun$ (model T2) and $\Mgas\simeq 7.6\times 10^7\Msun$ (model
T3), much less than the total DM mass $\MDM=10^9$. However, the gas
surface density drops abruptly at large radii: most ($\approx 90\%$
for model T2, $\approx80\%$ for model T3) of the gas mass lies within
$\rs$, while only 13\% of the DM mass ($1.3\times10^8\Msun$) is
contained within $\rs$ (Fig. \ref{fig:all}f).

For self-consistency we must have $\kinst \Rinst \gg1$
(i.e. $\lambdainst/2\pi \Rinst\ll 1$) and $\lambdainst/\hzinst\gg 1$,
where $\hzinst\equiv \hz(\Rinst)$ is the vertical scale height of the
disc at $\Rinst$.  The wavelength $\lambdainst$ of the fastest growing
unstable mode at this radius is reported in Table~\ref{tab:par}: we
note that $\lambdainst/2\pi\Rinst\lesssim0.3$ and
$\lambdainst/\hzinst\gtrsim7.3$, so the short-wavelength and thin-disc
approximations, on which the stability analysis is based, are
justified.  The mass of the fastest-growing unstable mode (equation
\ref{eq:Minst}) is $\Minst\approx 1.54\times10^5\Msun$ for model T2
and $\Minst\approx 6.05\times10^6\Msun$ for model T3.

\section{Coevolution of baryons and dark matter}
\label{sec:evol}

\subsection{Time-scales}
\label{sec:time}

The results of cosmological $N$-body simulations suggest that before
gas infall the DM halo is cuspy \citep{Nav95}. During gas infall the
halo tends to contract \citep{Blu86}, but the effect is small as long
as the gas density is lower than the DM density.  When the gas density
becomes comparable to or larger than the DM density, it is typically
high enough to trigger Toomre's instability, so the gas disc starts to
fragment into clumps.  The mass of these clumps is $\Mclump\approx
\Minst$, where $\Minst$ is the mass associated with the fastest
growing unstable mode (equation~\ref{eq:Minst}).

When orbiting through the distribution of the much lighter DM
particles, the massive clumps experience dynamical friction and
therefore tend to spiral in towards the halo centre.  The distribution
of DM particles is heated by the energy transferred from the clumps,
so the initial cosmological cusp of the DM halo is flattened by the
clumps \citep[e.g.][]{Elz01}. This process must happen on the
dynamical friction time-scale, which, for a clump of mass $\Mclump$ in
circular orbit at radius $r$ is \citep[][section 8.1]{BT08}
\begin{equation}
\tfric\simeq \frac{1.17}{\ln\Lambda}\frac{\MDM(r)}{\Mclump}\tcross,
\end{equation}
where $\MDM(r)$ is the DM mass contained within $r$, $\ln\Lambda$ is
the Coulomb logarithm and $\tcross\equiv r/\vcirc$ is the crossing
time. Applying the above formula to our reference models, fixing for
instance $r=\Rinst$, we get $\tfric/\tcross\simeq 33 /\ln\Lambda$ for
model T2 (in which $\tcross\approx 17\Myr$ at $r=\Rinst$) and
$\tfric/\tcross\simeq 1.4 /\ln\Lambda$ for model T3 (in which
$\tcross\approx 21\Myr$ at $r=\Rinst$). This suggests that, especially
in the higher-temperature case, dynamical friction heating acts on a
very short time-scale.

%%%%%%%%%%%%%%FIG 3
\begin{figure*}
\centerline{\psfig{file=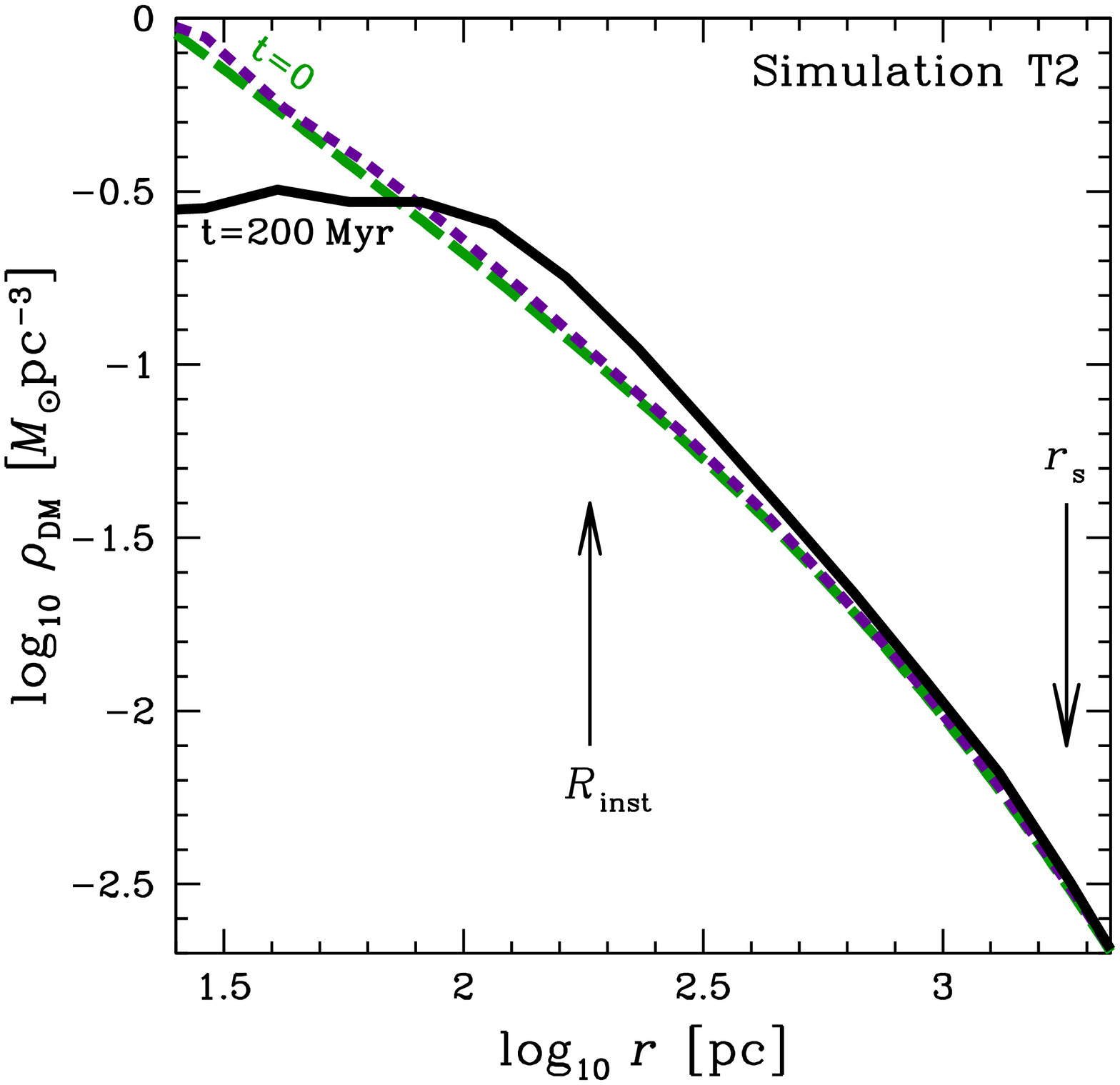,width=0.5\hsize}\psfig{file=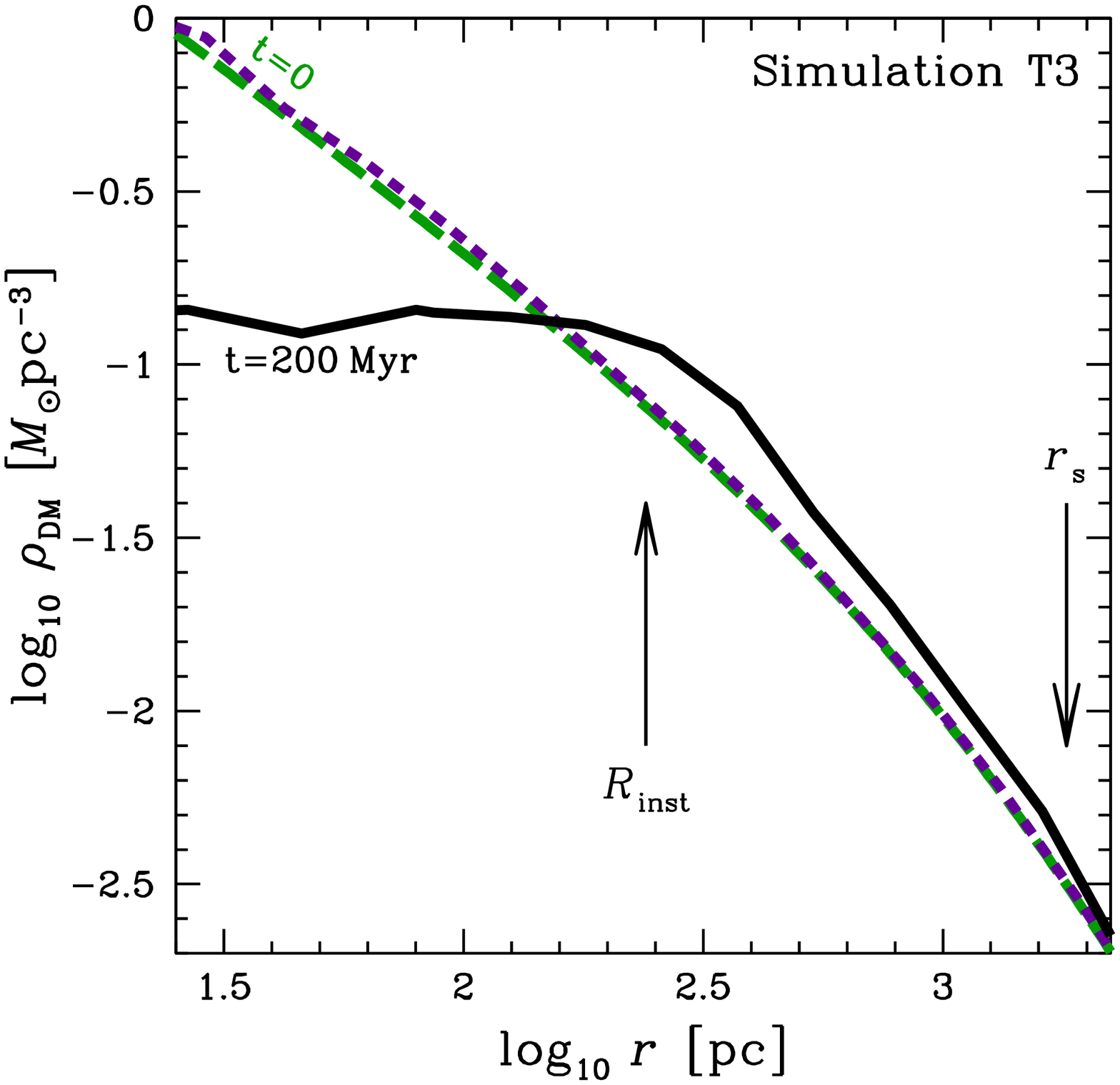,width=0.5\hsize}}
\caption{Initial ($t=0$; long-dashed curves) and final ($t=200\Myr$;
  solid curves) DM density profile in the presence of $\Nclump$ gas
  clumps of mass $\Mclump$ for models T2 (left-hand panel) and T3
  (right-hand panel) as predicted by $N$-body simulations.  In each
  panel the short-dashed curve is the profile obtained by letting the
  halo evolve in the absence of the clumps for $200\Myr$.}
\label{fig:den}
\end{figure*}
%%%%%%%%%%%%%%%%%%%%%%%

\subsection{$N$-body simulations}

To explore in detail the interaction between clumps and DM we ran $N$-body
simulations.  In each of these simulations the initial conditions consist of
$\Nclump$ clumps of mass $\Mclump$ and softening length $\varepsilonclump$ on
circular orbits in a spherical DM halo made of $\NDM=2\times 10^6$ particles.
In detail, the halo is represented by a truncated NFW distribution
\begin{equation}
\rhoDM (r)=\Mzero \frac{\exp\left[ -\left(r/\rt\right)^2\right]}{r \left(r+\rs\right)^{2}},
\end{equation}
where $\rs=1.81\kpc$, $\rt=1.23\rvir\simeq 12.67 \kpc$, and
$\Mzero\simeq9.6\times 10^8\Msun$: the values of these parameters are
such that the central density profile coincides with that of the model
described in Section~\ref{sec:dm} and the total DM mass is
$\MDM=10^9\Msun$.  We note that the DM mass within $300\pc$ is
$1.1\times10^7\Msun$, consistent with observations of many dSphs
\citep{Str08}.  Eddington's inversion \citep{BT08} is used to build
the halo in equilibrium in its gravitational potential with isotropic
velocity distribution.  The gravitational potential of the clumps is
neglected in setting up the halo: with this choice we implicitly
include also the effect of contraction of the halo due to the presence
of the baryons.

We present here results for two $N$-body simulations: simulation T2
with $\Mclump=1.54\times 10^5\Msun$, $\Nclump=76$ and
$\varepsilonclump=63\pc$ (based on model T2 in Section~\ref{sec:marg}),
and simulation T3 with $\Mclump=6.05\times10^6\Msun$, $\Nclump=12$ and
$\varepsilonclump=235\pc$ (based on model T3 in
Section~\ref{sec:marg}).  These values are such that $\Mclump\simeq
\Minst$, $\varepsilonclump\simeq \lambdainst/2$ and
$\Nclump\Mclump\approx\Mgas$ (see Table~\ref{tab:par}).  In the initial
conditions all the clumps move tangentially within the equatorial plane
$z=0$. The initial radial distribution of the clumps follows the
cumulative distribution function $f(R)=\Mgas(R)/\Mgas$ (see
Fig.~\ref{fig:all}f). In practice, we first distribute $n=\Nclump/2$
clumps at radii $R_i$ ($i=1,\ldots,n$) such that
$f(R_i)=\left(i-\frac{1}{2}\right)/n$, we extract the azimuthal angle
$\phi_i$ from a uniform distribution in $[0,\pi]$ and we assign
$\vR_i=0$ and $\vphi_i=\sqrt{G M_i/R_i}$, where
$M_i\equiv\MDM(R_i)+2(i-1)\Mclump$, where $\MDM(R_i)$ is the DM mass
within $r=R_i$. The other $n$ clumps are then given phase-space
coordinates $R_i=R_{i-n}$, $\phi_i=\phi_{i-n}+\pi$,
$\vphi_i=\vphi_{i-n}$ and $\vR_i=0$ ($i=n+1,\ldots,\Nclump$), so that,
even in the presence of the clumps, the centre of mass of the system is
at rest in the centre of the DM distribution.  The $N$-body
simulations were run with the parallel collisionless code \FVFPS
\citep[][]{Lon03,Nip03}. We adopted the following values of the code
parameters: minimum value of the opening parameter $\theta_{\rm
  min}=0.5$, softening parameter for the DM particles $\varepsilon =
0.02 \rs= 36 \pc$ and time-step $\Delta t=1.08\Myr$.

In both simulations T2 and T3 the effect of the presence of the clumps
is to flatten the DM cusp and to produce a central core. The formation
of the core is completed in $\approx 200\Myr$ in simulation T2 and
$\approx 150\Myr$ in simulation T3, so we take $t=200\Myr$ as
reference final snapshot.  Figure~\ref{fig:den} shows the DM profiles
of the two simulations at $t=0$ and $t= 200\Myr$.  A comparison with
the evolution of the halo in the absence of the clumps demonstrates
that the core is not a numerical artifact (see Fig.~\ref{fig:den}). As
expected, the effect is stronger in model T3 (higher gas temperature,
more massive clumps) than in model T2 (lower gas temperature, less
massive clumps).  The final central logarithmic density slope
$\gamma\equiv -\d\ln \rhoDM/\d \ln r$ is consistent with $\gamma=0$ in
both models, but the size of the core is significantly larger in
simulation T3 ($\approx 300 \pc$) than in simulation T2 ($\approx 100
\pc$).

The final DM mass within $300\pc$ is $1.4\times10^7\Msun$ in
simulation T2 and $1.3\times10^7\Msun$ in simulation T3, still
consistent with the observational data \citep{Str08}. The fact that
the DM mass within $300\pc$ is higher than in the initial conditions
($1.1\times10^7\Msun$) is a consequence of the contraction of the halo
due to the presence of the baryons.  As mentioned above, in our
experiments we implicitly account for the contraction of the halo
because the halo is set up in equilibrium in its own gravitational
potential, neglecting the potential of the clumps. The final profiles
shown in Fig.~\ref{fig:den} result from the combination of contraction
due to baryonic infall and expansion due to dynamical friction
heating.  We verified with simple $N$-body experiments that if the
same halo evolves in the absence of the clumps, but in the presence of
a fixed gravitational potential representing the corresponding smooth
gas distribution, the central DM profile quickly ($\lesssim 50\Myr$)
settles into a new equilibrium configuration with slightly steeper
central slope ($\gamma\simeq 1.1$ for model T2 and $\gamma\simeq 1.2$
for model T3) than the initial $\gamma=1$.  However, when the gas
fragments into clumps, the slight steepening due to contraction is
overwhelmed by heating through dynamical friction, and the final
profiles are cored with $\gamma\approx 0$.

\subsection{Lifetime of the gas clumps}
\label{sec:life}

For the mechanism described above to be effective, the lifetime of the
gas clumps must be at least of the same order as the time ($\sim
10^8\yr$) necessary to redistribute the DM in the central regions of
the halo.  In general, the question of the evolution of massive clouds
in galaxies is complex and highly debated, especially because it is
strictly related to the poorly understood process of star formation
and associated feedback \citep*[e.g.][]{Mur10,Gen12,Hop12,Bou14}.
However, in the specific context here considered, a few simple
arguments suggest that the assumption of a long clump lifetime is a
reasonable one.

It is widely accepted that any mechanism responsible for cloud
disruption starts to act after the formation of the first stars. At
$T\lesssim 10^4\K$ metal-poor gas has difficulty cooling, especially
in the absence of molecules, and molecules form with difficulty in the
absence of dust.  The time-scale for star formation cannot be shorter
than the longer of the cloud's initial dynamical time
$\tdyn=(G\rhogas)^{-1/2}$ and the cooling time $\tcool=3\kb T/2\L
\fHtwo$, where $\L=\L(T,\rhogas)$ is the cooling rate per molecule and
$\fHtwo=\nHtwo/\ngas$ is the molecular-hydrogen fraction ($\nHtwo$ is
the molecular-hydrogen number density and $\ngas=\rhogas/\mu\mp$ is
the total gas number density). Taking as gas density
$\rhogas(R=\Rinst,z=0)$ and estimating $\L$ from equation 2.32 of
\citet{Sti09}, for our models we get $\tdyn\simeq3.4\times10^7\yr$ and
$\tcool\simeq2.8\times10^9(\fHtwo/10^{-5})^{-1}\yr$ when
$T=3\times10^2\K$, and $\tdyn\simeq3.9\times10^7\yr$ and
$\tcool\simeq1.5\times10^7(\fHtwo/10^{-5})^{-1}\yr$ when
$T=3\times10^3\K$.  Notwithstanding the dependence of $\tcool$ on the
poorly constrained parameter $\fHtwo$, these numbers suggest that the
star-formation time-scale should not be shorter than a few tens of
megayears. Even when the first burst of star formation occurs, the
cloud is not immediately shattered: if cloud disruption is due to
supernovae, which however is controversial \citep{Mur10}, $\sim 10^7
\yr$ from the onset of star formation is required for massive stars to
complete their lives.

Independent arguments in favour of long cloud lifetimes derive from
observations of the Milky Way. Even the dusty, metal-enriched gas of
the current Galactic disc converts to stars extremely slowly in the
sense that less than one percent of a molecular cloud is turned into
stars in a dynamical time.  Moreover, since the overwhelming majority
of the gas interior to the Sun is in molecular form, molecular clouds
cannot be easily dissociated by star formation: their estimated
lifetime is $\gtrsim 10^8\yr$ \citep[][]{Sco13}.

Based on the above discussion, we expect clouds to orbit within the
dark halo for several dynamical times before they are dissipated by
stellar feedback. Thus we expect clouds to heat the halo for many
dynamical times and to leave a DM-dominated galaxy when they are
finally disrupted. However, it must be stressed that the lifetime of
the clumps is a crucial parameter of the presented model and it would
be highly desirable to obtain better constraints on it.

\section{Relation to previous work}
\label{sec:rel}

In this paper we have shown that, before star formation, dwarf-galaxy
sized cosmological halos are expected to host gaseous clumps with
masses, sizes and orbits such that they quickly flatten the original
DM cusp.  The idea that massive clouds can effectively heat stellar
systems dates back at least to \citet{Spi51}.  \cite{Elz01} proposed
that a similar mechanism (baryonic clumps heating DM halos) could be
responsible for turning DM cusps into cores. Subsequent work
elaborated on this proposal, by considering specific applications to
clusters of galaxies \citep{Elz04,Nip04} or exploring the space of
parameters by varying the density and velocity distribution of the
host system and the clump mass, size and orbits
\citep{Are07,Jar09,Goe10,Col11}. The results of these works clearly
indicated, as a proof of concept, that dynamical friction heating by
clumps can have an important role in halos on different scales and
that, quantitatively, the relevance of this process depends on the
properties of the clumps and of the host halo. However, not so much
work has been done to quantify this effect specifically on the scale
of dwarf galaxies. Here, though in an idealized framework, we try to
estimate self-consistently the masses, sizes and orbits of the clumps
and their effect on the DM distribution in a dwarf-sized halo of mass
$\sim 10^9\Msun$.  In this sense, among previous works, our
investigation is most similar to that of \citet{Ino11} who simulated,
with a Smoothed Particle Hydrodynamics code, the formation of a disc
galaxy in a relatively massive DM halo of $5\times 10^{11}\Msun$,
finding significant evolution of the halo density profile due to
interaction with baryonic clumps. However, these results cannot be
extrapolated to the smaller scale of dwarf galaxies, because, as
stressed above, a crucial quantity in the process is the ratio between
the gas temperature at the time of fragmentation (set by quantum
physics) and the virial temperature of the halo (set by gravity).

As an alternative to the mechanism considered in the present work, it has
been suggested that the dominant factor in determining the final DM
distribution is fluctuations of the gravitational potential that are driven
by stellar feedback \citep[e.g.][]{Mas06,Mas08}.  In such a scenario energy
is transferred to the DM by the bulk motions of gas driven by repeated bursts
of star formation.  Though this scenario shares some features with the one
considered in the present paper (gas clumps moving in the halo transfer
energy to the DM), it must be stressed that the two scenarios are
conceptually and materially different\footnote{The simulations of
\citet{Mas06} are only superficially
  similar to those presented in Section~\ref{sec:evol}: in their
  model, aimed at mimicking the gas bulk motions induced by successive
  episodes of stellar feedback, three massive clumps move as harmonic
  oscillators through the system's centre.}: in one case the process
occurs before star formation and the energy source is gravitational,
in the other case the process occurs after star formation and the
energy source is stellar feedback.  The two mechanisms might be
alternative or even cooperate during the formation and evolution of
dSphs (see, for a discussion, \citealt{Pon14}). Remarkably, if the
mechanism studied in our work is effective, it occurs before star
formation and therefore earlier with respect to stellar-feedback
driven processes. The effectiveness of the dynamical friction heating
mechanism depends crucially on the lifetime of the clumps, which is a
debated question.  In Section~\ref{sec:evol} we have argued that, at
least in the considered case of dwarf galaxies, the primordial clumps
are expected to live long enough to erase the central DM cusp.

\section{Conclusions}
\label{sec:con}

Dwarf spheroidal galaxies are now DM dominated, but stars cannot form
until gas becomes locally gravitationally dominant. Dissipation causes
gas to move inward within a DM halo until this condition is
satisfied. When the mass of gas is small compared to the DM mass,
considerable contraction is required to achieve gravitational
dominance, and near-conservation of angular momentum will cause the
gas distribution to become flattened. A differentially rotating body
of gas will be heated by the processes that heat accretion discs, so
its temperature will be much greater than that of the cosmic
background radiation even in the case in which heating at the time of
infall is extremely inefficient. In the absence of cooling by metal
ions, it seems likely that, the temperature of the gas will rise to
temperatures $\sim3000\K$ at which cooling by hydrogen becomes
non-negligible, but we have considered also temperatures an
order-of-magnitude lower.

In the potential of a low-mass halo, the disc may not be very thin,
but equilibrium dynamical models of bodies of spinning gas moving in
the common gravitational potential of the gas and DM suggest that it
is probably thin enough for Toomre's seminal analysis of the stability
of thin discs to yield the wavelength at which fragmentation first
occurs to within a factor of a few. This wavelength in turn specifies
the mass of the fragments into which the gas divides. The key point is
that this mass is not very much smaller than the mass of DM interior
to the radius where fragmentation starts.  Consequently, only a few
local dynamical times are required for the fragments to transfer
significant energy to the local DM particles.  {Therefore the
  halo's central cusp is heated to form a core with central
  logarithmic density slope $\gamma\approx 0$ before stellar feedback
  enters the scene.}

We have simulated this transfer with an $N$-body code, and find that
it quickly erases the cusp at the centre of the original DM halo.
This result is not at all surprising; the existence of DM cusps has
always been rather puzzling since they depend on the velocity
dispersion within the halo decreasing to zero as one approaches the
centre. Any gravitational scattering of the ultra-cool particles that
form the cusp will raise the central velocity dispersion and erase the
cusp. {Hence the wonder is that such scattering does not occur
  during hierarchical merging of DM}.

In a DM only simulation, the only objects capable of such scattering
within a given halo are halos of smaller mass that reach the centre of
the host halo. These halos have at their centres cusps that are dense,
so are not tidally disrupted until the satellite halo has reached the
very centre of it host \citep[e.g.][section 9.3.2]{BT08}. It seems
likely that cusp particles in the host halo that are upscattered by
the satellite halo are efficiently replaced by the satellite halo's
own cusp particles.

Gas can easily form clouds massive enough to scatter particles out of
the host halo's cusp, and a gas cloud does not replace these
particles. When the massive stars that form within the cloud disrupt
it, the host halo is left DM dominated and in possession of a core.

This cusp erasing process will be efficient unless the gas cools to
such low temperatures that it does not fragment until a {\it very\/}
thin disc has formed. Since quantum physics sets the minimum
temperature for efficient cooling, discs can be very thin at the onset
of fragmentation only in higher-mass halos (which have higher virial
temperatures). In these halos star formation is known to be more
efficient in the sense that a greater fraction of gas is converted to
stars before residual gas is returned to the intergalactic medium. In
this case the stellar disc is likely to be self-gravitating, so it can
become bar-unstable.  A stellar bar quickly transfers energy to a
co-spatial DM cusp \citep{Tre84,Sel08}, so spiral galaxies with bars
will not have DM cusps either. In principle there might be an
intermediate halo mass at which star formation is so inefficient that
it leaves a completely DM dominated stellar disc, and yet at the onset
of fragmentation the gas disc is sufficiently thin for the resulting
gas clouds to be inefficient scatterers. Then in these
intermediate-mass halos DM cusps would survive star formation. But the
existence of this intermediate range of halo masses seems unlikely, so
stars and DM cusps are probably mutually exclusive.

\section*{Acknowledgements}
CN acknowledges financial support from PRIN MIUR 2010-2011, project
``The Chemical and Dynamical Evolution of the Milky Way and Local
Group Galaxies'', prot. 2010LY5N2T.  JB was supported by STFC by
grants R22138/GA001 and ST/K00106X/1. The research leading to these
results has received funding from the European Research Council under
the European Union's Seventh Framework Programme (FP7/2007-2013) / ERC
grant agreement no.\ 321067.

\end{document}